# A Multi-Technique Study of $C_2H_4$ Adsorption on $Fe_3O_4$(001)

Lena Puntscher[1#], Panukorn Sombut[1#], Chunlei Wang[1], Manuel Ulreich[1], Jiri Pavelec[1], Ali Rafsanjani-Abbasi[1], Matthias Meier[1,2], Adam Lagin[1], Martin Setvin[1,3], Ulrike Diebold[1], Cesare Franchini[2,4], Michael Schmid[1] and Gareth S. Parkinson[1*]

[1]Institute of Applied Physics, TU Wien, Vienna, Austria
[2]Faculty of Physics, Center for Computational Materials Science, University of Vienna, Vienna, Austria
[3]Department of Surface and Plasma Science, Faculty of Mathematics and Physics, Charles University, Prague, Czech Republic
[4]Dipartimento di Fisica e Astronomia, Università di Bologna, Bologna, Italy

[#]These authors contributed equally to this work

ABSTRACT

The adsorption/desorption of ethene ($C_2H_4$), also commonly known as ethylene, on $Fe_3O_4$(001) was studied under ultrahigh vacuum conditions using temperature programmed desorption (TPD), scanning tunneling microscopy, x-ray photoelectron spectroscopy, and density functional theory (DFT) based computations. To interpret the TPD data, we have employed a new analysis method based on equilibrium thermodynamics. $C_2H_4$ adsorbs intact at all coverages and interacts most strongly with surface defects such as antiphase domain boundaries and Fe adatoms. On the regular surface, $C_2H_4$ binds atop surface Fe sites up to a coverage of 2 molecules per ($\sqrt{2}\times\sqrt{2}$)R45° unit cell, with every second Fe occupied. A desorption energy of 0.36 eV is determined by analysis of the TPD spectra at this coverage, which is approximately 0.1–0.2 eV lower than the value calculated by DFT+U with van der Waals corrections. Additional molecules are accommodated in between the Fe rows. These are stabilized by attractive interactions with the molecules adsorbed at Fe sites. The total capacity of the surface for $C_2H_4$ adsorption is found to be close to 4 molecules per ($\sqrt{2}\times\sqrt{2}$)R45° unit cell.

1.   Introduction

Iron oxides are some of the most abundant materials on earth. Their primary usage is as a feedstock for the steel industry, but their low toxicity, natural abundance, and magnetic properties make them popular in many applications. In catalysis, iron oxides serve as an active component for processes such as Fischer-Tropsch synthesis and the water-gas shift reaction[1-3]. They are also frequently utilized as a robust, inexpensive, reducible support for precious metal nanoparticle catalysts.

In recent years the field of "single-atom catalysis" (SAC) has emerged as an intensely studied topic in catalysis research, and iron oxides continue to be a common choice of support material[4-10]. Such catalysts are typically prepared by co-precipitation, and the support is nominally an $Fe_2O_3$ powder. Nevertheless, it is often labelled $FeO_x$ to reflect that the oxide, and especially its surface, is reduced when the catalyst is activated by heating in CO or $H_2$. In recent years, we have utilized $Fe_3O_4(001)$ as a model support to study fundamental processes in single-atom catalysis. This work is enabled by a ($\sqrt{2}\times\sqrt{2}$)R45° surface reconstruction, based on an ordered array of subsurface interstitials and vacancies[11], which stabilizes adsorbed transition metal cations up to temperatures as high as 700 K[12-17].

To date, one of the major applications of SAC has been the hydrogenation of alkenes[18, 19]. There is also evidence that SAC can selectively catalyze the hydroformylation of alkenes to aldehydes[20, 21]. This reaction is usually catalyzed by coordination complexes in solution. The heterogenization of reactions currently performed by homogeneous catalysts is a particularly exciting target for SAC research[22-24]. However, prior to studying the role of single atoms in catalyzing complex multi-reactant processes, it is important to understand how the individual reactants interact with the support.

In this paper, we study how the simplest alkene, $C_2H_4$, interacts with the $Fe_3O_4(001)$ surface. This work follows up on a recent study by Lee et al.,[25] who performed TPD and XPS measurements of $C_2H_2$, $C_2H_4$ and $C_2H_6$ on $Fe_3O_4(001)$ and concluded that $C_2H_4$ physisorbs weakly (adsorption energies between 0.29 eV and 0.41 eV) with four desorption peaks within the first monolayer. We reproduce the TPD data for $C_2H_4$ here, with the addition of three additional peaks at low temperature that are accessible due to the lower adsorption temperature employed in our measurements. Moreover, we supplement these data with STM images and DFT+U calculations and show how it is that the molecules are accommodated on the $Fe_3O_4(001)$ surface. Specifically, we show that $C_2H_4$ preferentially adsorbs at defects including anti-phase domain boundaries, and then atop surface $Fe^{3+}$ atoms up to a coverage of 2 molecules per ($\sqrt{2}\times\sqrt{2}$)R45° unit cell. Additional $C_2H_4$ are stabilized between the Fe rows through attractive intermolecular interactions up to a coverage of 4 $C_2H_4$ molecules per ($\sqrt{2}\times\sqrt{2}$)R45° unit cell.

## 2. Experimental and Computational Details

The experiments were performed on natural $Fe_3O_4(001)$ crystals (6×6×1 mm, SurfaceNet GmbH). The samples were prepared in ultrahigh vacuum (UHV) by cycles of sputtering followed by annealing at 923 K for 20 min. For the STM experiment we used 10 minutes of 1 keV $Ar^+$ sputtering with a target current of 0.8 µA. For the TPD/XPS experiments we used 10 minutes of 1 keV $Ne^+$ sputtering with a target current of 1 µA. After every other cycle, the sample was oxidised by exposure to $O_2$ during annealing ($2×10^{-6}$ mbar $O_2$, 20 min), which leads to the growth of a pristine $Fe_3O_4(001)$ surface by migration of interstitial Fe from the sample bulk to the surface[26].

Two separate UHV setups were used to carry out the experiments. The STM data were acquired in a two-vessel UHV chamber consisting of a preparation chamber (p < $10^{-10}$ mbar) and an analysis chamber (p < $2×10^{-11}$ mbar). The analysis chamber is equipped with a Tribus STM head (Sigma Surface Science) and a low-noise in-vacuum preamplifier[27]. The STM measurements were conducted in constant current mode with an electrochemically etched W tip. The STM images were corrected for distortion and creep of the piezo scanner as described in ref. 28. $C_2H_4$ (Messer, 99.95 %) was leaked into the analysis chamber at various pressures (up to a maximum of $5×10^{-9}$ mbar) through an open door in the thermal shield of the liquid-nitrogen-cooled STM head. The sample was held at 78 K. In experiments where a colder sample temperature was required, the nitrogen in the cryostat was pumped. This way a temperature of 68 K could be reached. The gas doses given for these STM experiments were measured in the analysis chamber; those at the sample are likely lower than the measured values. For adsorption, the tip was lifted to avoid shadowing the incoming molecules, thus images for different gas doses do not show the same position on the sample.

The TPD and XPS spectra were obtained in a second vacuum system optimized to study the surface chemistry of single-crystal metal-oxide samples. The samples are mounted on a Ta backplate using Ta clips, with a thin Au sheet in between to aid the thermal contact. The sample is cooled using a liquid-He flow cryostat, and can be counter heated to the dosing temperature (60 K) or higher temperatures via resistive heating of the Ta backplate[29]. The vacuum system is equipped with a home-built effusive molecular beam source based on an orifice with effective diameter 38.0 ±1.9 µm, which delivers a close to top-hat profile at the sample with a 3.5 mm diameter and a beam core pressure of 3.0 ±0.3 ×$10^{-8}$ mbar at the sample position[29, 30]. The base pressure in the chamber is below $10^{-10}$ mbar. A quadrupole mass spectrometer (Hiden HAL 3F PIC) is used in a line-of-sight geometry for TPD experiments, and a monochromatized Al/Ag twin anode X-ray source (Specs XR50 M, FOCUS 500) and a hemispherical analyzer (Specs Phoibos 150) are used for XPS measurements. The energy scale is calibrated after each bakeout using copper, silver and gold foils attached to the cryostat. A complete description of the chamber design and capabilities is given in ref. 29.

The Vienna *ab initio* Simulation Package (VASP) was used for all DFT calculations[31, 32]. We adopted the strongly constrained and appropriately normed meta-generalized gradient approximation (SCAN) [33] with the inclusion of van der Waals interactions (rVV10)[34] and an on-site Coulomb repulsion term $U_{eff}$=3.61 eV[35, 36] for Fe atoms to model the oxide. The surface calculations are based on the subsurface cation vacancy (SCV) reconstructed model of the $Fe_3O_4$(001) surface[11], using the Γ-point only for the (2√2 × 2√2)R45° supercell. Calculations were performed with the experimental magnetite lattice parameter (a=8.396 Å) using an asymmetric slab with 13 planes (7 planes with octahedral Fe and 6 with tetrahedral Fe; the bottom 9 planes are fixed and only the 4 topmost planes relaxed) and 14 Å vacuum. Convergence is achieved when an electronic energy step of $10^{-6}$ eV is obtained, and forces acting on ions smaller than 0.02 eV/Å, with the plane-wave basis cutoff energy set to 550 eV. Note that the SCV reconstruction is oxidized with respect to bulk $Fe_3O_4$, and that all Fe in the outermost 4 layers are $Fe^{3+}$ like. Consequently, there is a small bandgap in the surface layers. The $Fe_3O_4$ bulk also exhibits a small bandgap in our setup, and thus represents the sub-Verwey transition (<125 K) phase[37].

The average adsorption energy per $C_2H_4$ molecule is computed according to the formula

$$E_{ads,a} = \left(E_{Fe_3O_4+nC_2H_4} - \left(E_{Fe_3O_4} + nE_{C_2H_4}\right)\right)/n \qquad (1)$$

where $E_{Fe_3O_4+nC_2H_4}$ is the total energy of the $Fe_3O_4$(001) surface with adsorbed $C_2H_4$, $E_{Fe_3O_4}$ is the total energy of the clean $Fe_3O_4$(001) surface, the $E_{C_2H_4}$ represents the energy of the $C_2H_4$ molecule in the gas phase, and $n$ is the number of $C_2H_4$ molecules.

The average adsorption energy corresponds to the stability of the system and is used as a search criterion to determine the lowest-energy configuration for a given coverage. However, the average adsorption energy is not what is observed experimentally in a TPD experiment. Rather, the peaks in TPD correspond to the energy required to remove the most weakly bound molecule from a given configuration, i.e., the differential adsorption energy. We define the differential adsorption energy accordingly to equation (2):

$$E_{ads,d} = E_{Fe_3O_4+nC_2H_4} - \left(E_{Fe_3O_4+(n-1)C_2H_4} + E_{C_2H_4}\right) \qquad (2)$$

where we assume the system starts with the lowest-energy configuration for $n$ molecules, and the non-desorbing molecules are able to freely relax and reach the new $n-1$ lowest-energy configuration without hindrance (such as barriers). This assumption should be fulfilled, since the rearrangement of the remaining molecules is limited to rotations and minor relaxations; no major rearrangement is necessary. The calculated C 1s core-level binding energies are calculated in the final state approximation[38].

## 3. Results

### 3.1 Temperature Programmed Desorption

Figure 1 shows a series of TPD spectra for various initial coverages of 0.3–12.7 $C_2H_4$ molecules per ($\sqrt{2}\times\sqrt{2}$)R45° unit cell. The absolute coverages were determined using the known flux of the molecular beam, the dosing time, and the experimentally determined coverage-dependent sticking coefficient determined using the King and Wells method[39]. The sticking probability is initially 0.975 at 60 K, but increases quickly to unity as molecules accumulate on the surface (see Fig. S1). In total, 7 desorption features are observed due to desorption of molecular $C_2H_4$ from $Fe_3O_4$(001). The two peaks in the range 70-75 K, labelled α and β in Fig. 1, are attributable to $C_2H_4$ multilayer desorption. The peaks at higher temperatures, labelled γ-η, originate from the first monolayer. We also observe a desorption signal at 215 K that is already present for zero nominal dose, and does not saturate with increasing coverage. Tests showed that $C_2H_4$ desorbs from the Ta backplate at this temperature, so we conclude this to be an experimental artifact originating primarily from adsorption of $C_2H_4$ outside the beam spot due to a slight increase of the $C_2H_4$ partial pressure during the TPD series. Consequently, we subtracted the spectrum for zero dose from all other datasets shown in Fig. 1. Overall, our TPD data for the $C_2H_4$ system resemble those published recently by Lee et al.[25] for the same system, although the α, β and γ peaks are not visible in Ref. 25 because the dosing temperature (≈ 80 K) was higher than employed here. Also, the shoulder we observe at 125-135 K (ζ) is a clear peak in the dataset acquired by Lee et al.[25]. In what follows we discuss the TPD peaks in descending temperature order.

The TPD data for the lowest exposure performed (0.3 molecules per unit cell, which corresponds to 0.075 molecules per surface Fe atom) already contains the η peak at ≈160 K, and the onset of the ζ shoulder. Given the low coverage, these peaks must correspond to defects on the $Fe_3O_4$(001) surface. The η peak exhibits the typical behavior for a strongly binding defect site, saturating already at a low coverage and remaining at this intensity as the coverage is increased. The ζ shoulder around 130 K is peculiar for two reasons. First, it exhibits different intensities for different samples. In Figure S2, we show a TPD curve acquired on a different sample in our setup, where this peak is significantly larger, and the data more closely resembles that shown by Lee et al.[25]. Second, as already noted by Lee et al.[25], this feature does not completely saturate prior to the onset of the ε peak. We will revisit the likely origin of the ζ peak in the discussion section.

The ε peak emerges at 115-120 K and shifts to lower temperatures with increasing coverage, eventually saturating with a maximum at 110-115 K for a coverage of 2.2 (±0.2) $C_2H_4$ per unit cell. Next, two very sharp peaks, δ and γ, emerge, which saturate close to 3.0 (±0.3) and 4.1 (±0.4) $C_2H_4$ per unit cell, respectively. As the coverage increases there is non-zero desorption rate which rapidly shifts to lower coverage (typical for a compression close to monolayer coverage), before the onset of the multilayer peaks. The multilayer region contains two peaks, α and β. The β peak emerges first (blue curves in the inset in Fig. 1) and saturates at a coverage of 8.4 (±0.8) $C_2H_4$ molecules per unit cell (equivalent to approximately 2 layers of ethylene). For higher coverages it is replaced by the α peak (red curves in the inset of Fig. 1), which grows in intensity as the coverage is increased. Both multilayer peaks, α and β, exhibit the regular zero-order profile typical of multilayer desorption.

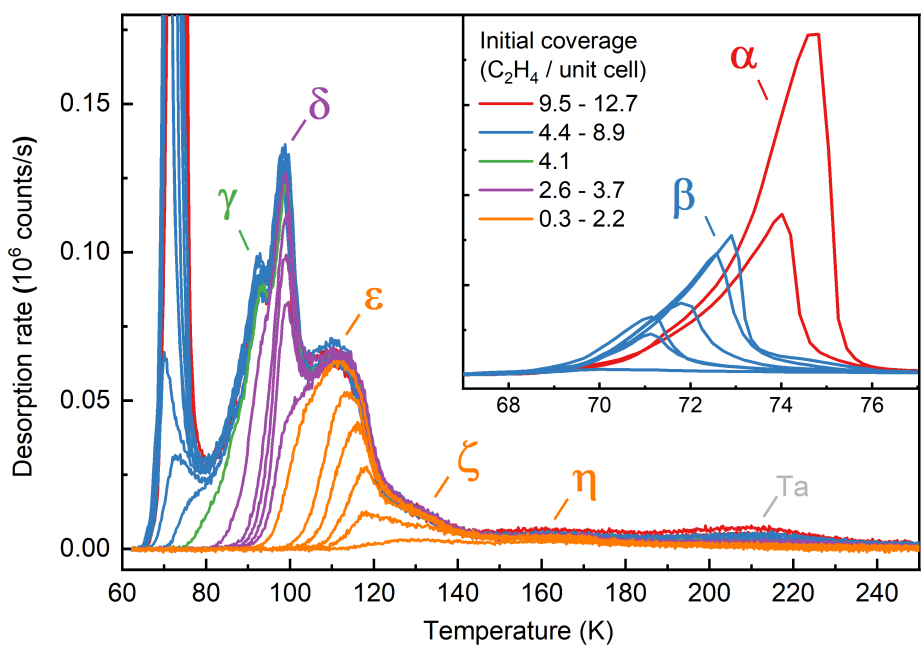

**Figure 1: TPD spectra for coverages ranging from 0 to 12.7 $C_2H_4$ molecules per $Fe_3O_4$(001) surface unit cell.** $C_2H_4$ was dosed at 60 K using a molecular beam and the TPD spectra were acquired using a temperature ramp of 1 K/s. Seven desorption features are labelled α-η. Peaks α and β result from desorption from multilayer $C_2H_4$, while γ, δ and ε originate from the first monolayer. The ζ and η peaks are due to adsorption at surface defects. The desorption feature at 215 K is an experimental artifact caused by adsorption of $C_2H_4$ on the Ta sample plate.

To further interpret the TPD data, we have employed a new analysis program fully described in Ref. 40. The procedure is based on equilibrium thermodynamics, and builds on the approach pioneered by Kreuzer et al.[41]. Note that this method does not attempt to determine a pre-exponential factor from experiment, but rather calculates the relevant quantities from thermodynamics. This analysis method is based on a *gedankenexperiment* where one holds the TPD ramp at any temperature and supplies a gas with the right pressure to establish adsorption-desorption equilibrium. Then, at this temperature, the chemical potential of the (hypothetical) gas phase and that of the adsorbate (at the given coverage) are equal. The chemical potential of the gas phase can be easily calculated, that of the adsorbates depends on the adsorption energy and the configurational and vibrational entropy of the adsorbate. This allows us to determine the adsorption energies.

In our TPD analysis, the adsorbates are treated as a lattice gas (diffusion barrier well above $k_B T$) on a surface containing a distribution of different adsorption sites. As the analysis is based on equilibrium thermodynamics, it requires that adsorbate diffusion is much faster than desorption, to establish equilibrium between the adsorbates. This condition is usually fulfilled for nearby adsorption sites, but not necessarily for different surface areas with a large separation, i.e., large domains or crystallites with different structure. The analysis does not explicitly take interactions between adsorbates into account; nevertheless, short-range repulsion is modelled as the occupation of sites with steadily weaker adsorption energy as the coverage increases. Our model does not include attractive interactions. These are usually recognizable by the experimental TPD peaks being sharper than the simulated ones.

The energy and entropy of the adsorbates are influenced by the low-frequency vibrational modes, i.e., the hindered translation and rotation modes, which can be determined by DFT. Specifically, our calculations in the limit of low coverage yield $h\nu$ = 4.5, 7.8, 11.0, 14.0, 15.8, and 19.1 meV for these modes. Since the influence of the vibrations on the adsorption energies is only about 5%, we neglect the coverage dependence of the vibration frequencies. The results presented here assume Langmuirian sticking with an initial (low-coverage) sticking coefficient of $s_0 = 1$, but the calculated adsorption energies change by less than 3% if we instead assume either a coverage-independent sticking coefficient close to unity (as observed in our experiments during adsorption at low temperatures) or $s_0 = 0.3$. The adsorption energy distribution is obtained from the TPD curve at saturation coverage (i.e., the 4.1 (±0.4) molecules per unit cell), excluding the multilayer peaks.

Since the experimental TPD curves show a slight increase of the intensity with coverage at temperatures above 150 K, which we attribute to readsorption of desorbed gas from the sample on the Ta backplate, we have corrected the input for the TPD analysis program by subtracting this linear coverage dependence at high doses (apart from the noise of the experimental data, the black curve at $T$ > 150 K in Fig. 2b is identical to the corrected TPD spectrum).

The resulting adsorption energy distribution is shown in Fig. 2a, and coverage-dependent TPD curves calculated from this energy distribution are shown in Fig. 2b. The experimental background-corrected intensities also shown in this plot have been scaled with $\sqrt{T}$ to account for the velocity-dependent ionization probability[40]. We find good agreement between experiment and simulation for the ε peak. The discrepancies between experiment and calculation concerning the γ, δ, and ζ peaks are considered in the discussion.

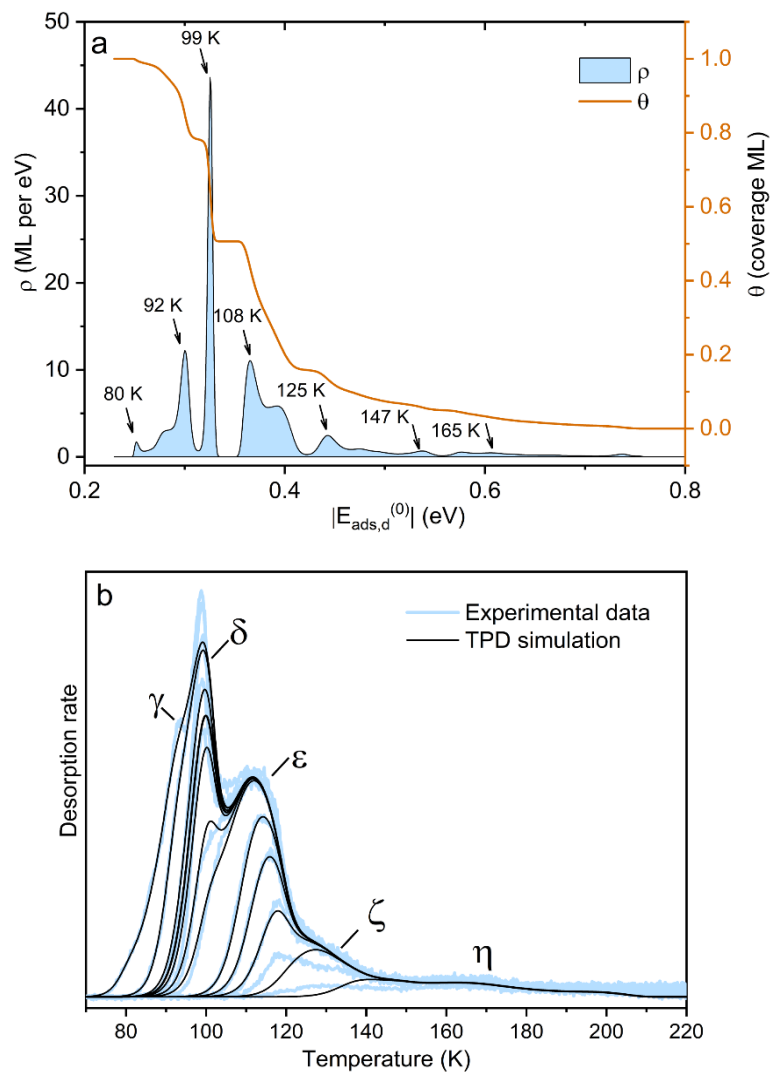

**Figure 2: Analysis of the TPD spectra** using the method presented in ref 40. (a) Distribution of the adsorption energies derived from the TPD trace at saturation of the first monolayer (ML). The brown curve shows the cumulative distribution function, i.e., the fraction of molecules with stronger adsorption than the given value on the x-axis. Temperatures indicated at the peaks are based on a theoretical correspondence between the chemical potential of the adsorbate and the desorption temperature[40] and do not exactly agree with the desorption peaks. (b) TPD spectra calculated from the adsorption energy ($E_{ads,d}$) distribution in (a), plotted on top of the experimental data (corrected for background and velocity-dependent ionization probability). The model does not include attractive interactions between adsorbates, leading to a poor fit of the γ and δ peaks. Possible reasons for the discrepancies between experiment and calculation in the region of the ζ shoulder are discussed in section 5.

## 3.2   Photoelectron Spectroscopy

XPS measurements were performed to gain information about the chemical state of the adsorbed molecules and the $Fe_3O_4(001)$ surface (Fig. 3).

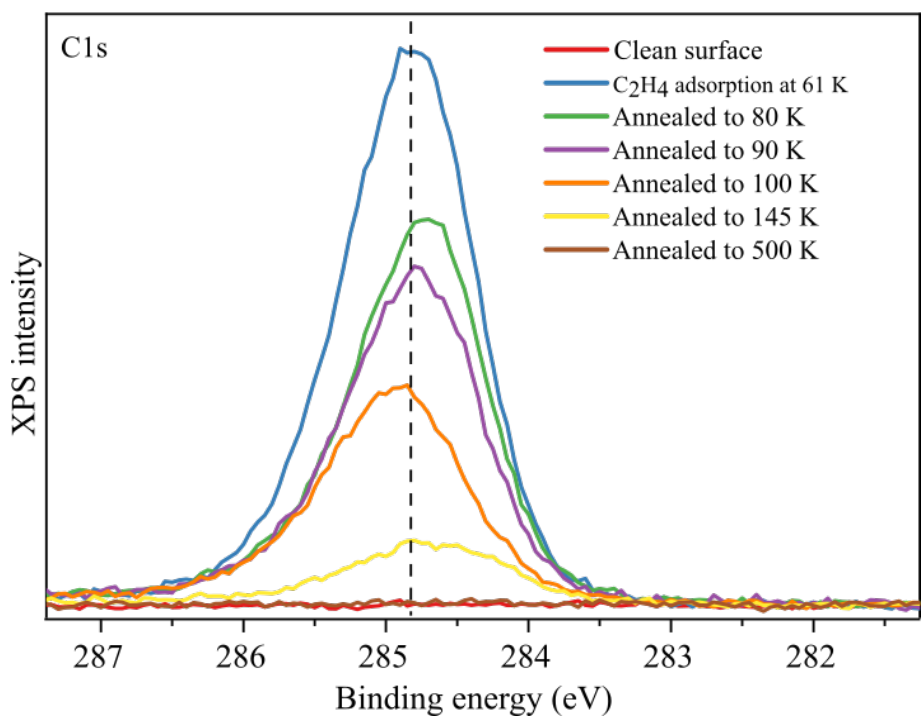

**Figure 3: X-ray photoelectron spectroscopy data for $C_2H_4/Fe_3O_4(001)$ measured at 61 K.** The pristine surface exhibits no detectable C1s signal. Following the adsorption of multilayer $C_2H_4$ (specifically 12.7 $C_2H_4$ per unit cell) at 61 K, a C1s peak is visible at 284.8 eV, a binding energy typical for C=C bonding. Annealing to 80 K removes the molecules from the multilayer (peaks α and β in Fig. 1). The C1s signal from the saturated first monolayer (green) is slightly shifted to lower binding energy (248.7 eV). Subsequent annealing steps sequentially remove $C_2H_4$ molecules from the monolayer until the sample is again free from C after annealing to 500 K. Spectra were acquired with Al Kα radiation in grazing emission (12° from normal).

XPS data were acquired for the pristine $Fe_3O_4(001)$ surface, after $C_2H_4$ adsorption, and after several heating steps up to 500 K. Fig. 3 shows the C 1s region. (Corresponding data from the Fe 2p and O 1s regions are shown as Fig. S3.) The as-prepared surface is free of C within the detection limit of the setup. A peak due to multilayer $C_2H_4$ appears at 284.8 eV after exposure to 12.7 $C_2H_4$ per surface unit cell at 61 K, with a second peak shifted by 8.3 eV to higher binding energy due to a π-3p Rydberg shake-up process (Fig. S4)[42-45]. After the sample was heated to 80 K, the area of the C1s peak decreases by 50%. Subsequent heating to 90 K and 100 K further reduces the intensity of the C1s peak as the molecules contained within the first monolayer desorb. The intensity after annealing to 100 K corresponds to approximately half of the complete monolayer, which makes sense as the molecules contained within the ε peak (≈2 $C_2H_4$ per surface unit cell) should remain on the surface at 100 K. Removing these molecules by heating to 145 K leaves only the molecules associated with surface defects. The fact that the binding energy remains unchanged shows that $C_2H_4$ is adsorbed molecularly even at defects. Finally, after heating to 500 K the sample is free from C within the detection limit of the instrument. XPS data from the Fe 2p and O 1s regions after $C_2H_4$ adsorption are representative of the clean $Fe_3O_4$(001) surface throughout the experiment, and show only a decrease in intensity when $C_2H_4$ is adsorbed (see Fig. S3). The peak positions indicate that the molecules corresponding for the ε peak (orange curve in Fig. 3) have a slightly higher binding energy (XPS peak shifted by 0.2 eV) than those completing the monolayer (γ and δ peaks) or those adsorbed at defects (yellow curve in Fig. S3).

## 3.3 Scanning-Tunneling Microscopy

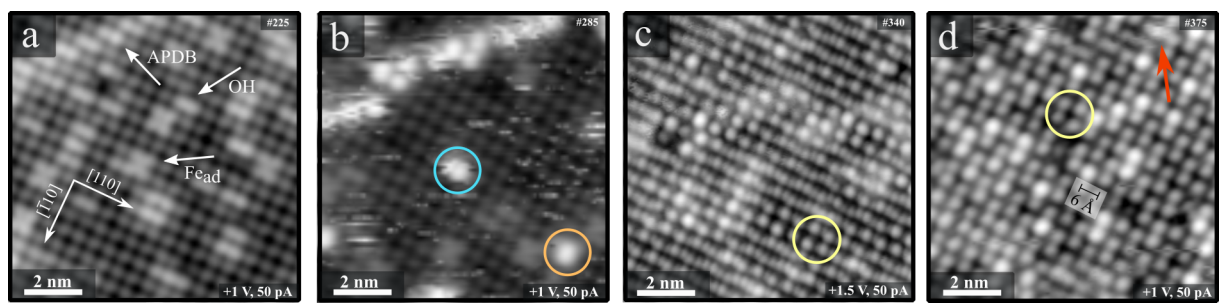

**Figure 4: Empty state STM images (10×10 nm²) of the $Fe_3O_4$(001) surface before and after exposure to $C_2H_4$.** Images (a–c) were obtained at $T$ = 78 K. (a) The as-prepared $Fe_3O_4$(001) surface exhibits rows of 5-fold coordinated Fe atoms along [110]. Each protrusion in the image corresponds to a pair of these Fe atoms[11]. Typical surface defects are labelled (see main text for details). (b) After exposure to 0.5 L of $C_2H_4$ at 78 K. Bright protrusions titrate only the surface defects. (c) Upon exposure to $C_2H_4$ (5 L, dosed at 5×10⁻⁹ mbar) the surface Fe rows remain visible, as well as bright circular protrusions that order locally with a 6 Å square lattice (e.g. yellow circle). (d) An image acquired after lowering the temperature to 68 K and exposing to more ethylene (2.3 L, at 5×10⁻⁹ mbar) exhibits a nearly full coverage of bright protrusions separated by 6 Å along the [110] direction. The red arrow marks indications of mobility of the $C_2H_4$ molecules.

Figure 4 shows STM images of the $Fe_3O_4(001)$ ($\sqrt{2} \times \sqrt{2}$)R45° surface before and after exposure to different amounts of $C_2H_4$. Liquid nitrogen was used as the cryogen for the low-temperature (LT)-STM, which results in a sample temperature of 78 K for images (a-c). For image (d), pumping of the cryostat lowered the sample temperature to 68 K. Figure 4a is a typical image of the as-prepared $Fe_3O_4(001)$ surface, showing the characteristic $Fe^{3+}$ atoms of the clean surface [11]. Three different surface defects appear as bright protrusions: iron adatoms ($Fe_{ad}$) [46], surface hydroxyls (OH) [47], and antiphase domain boundaries (APDB) [48]. After exposing the sample to 0.5 Langmuir (1 L = 1.33 mbar·s) at 78 K, bright protrusions appear at the defect sites (Fig. 4b). Note that since $C_2H_4$ was dosed into the cryostat through a small window, the exposure is significantly less than the nominal dose. Titration of the defect sites at higher temperatures is consistent with the TPD experiments, as desorption from surface defects is assigned to the ζ and η TPD peaks. Thus, diffusion must be sufficiently facile at 78 K for the molecules to locate the defects in the time frame of the experiment. A line of bright protrusions on the Fe rows in the top left of the image corresponds to $C_2H_4$ adsorbed on an APDB. The APDB has previously been shown to be an active site for $CH_3OH$ adsorption [46]. Two different adsorption sites are marked in the image: The cyan circle shows an isolated protrusion located above the Fe row, which corresponds to $C_2H_4$ adsorbed atop a five-fold coordinated surface Fe atom in a regular lattice position. This could be due to an unreconstructed unit cell defect [46], which occurs when an additional Fe atom is accommodated in the subsurface, or potentially next to an OH group [49]. However, many OH groups remain visible in Fig. 4b, so we infer that the defect responsible for $C_2H_4$ adsorption is probably the unreconstructed unit cell. The orange circle shows a bright protrusion between the surface Fe rows, which corresponds to $C_2H_4$ adsorption at a two-fold coordinated Fe adatom defect. Some streakiness present in the image is attributed to molecules moving on the surface during the STM measurement.

Exposing the sample at 78 K to a nominal exposure of 5 L of $C_2H_4$ (Fig. 4c) leads to bright protrusions on the surface Fe rows, similar to those observed in Fig 4b. In large areas of the image, the nearest-neighbor distance of the protrusions along the surface Fe row direction is 6 Å, which corresponds to every other Fe cation being occupied by a $C_2H_4$ molecule (yellow circle). The underlying Fe rows remain visible in patches, but the resolution is enhanced with clear protrusions at the position of all surface Fe atoms (nearest neighbor distance = 3 Å, different from the usual appearance where pairs of Fe atoms appear as protrusions, with a distance of 6 Å along the rows). There appears to be a gradual transition between the areas where every other Fe position appears bright due to an adsorbed $C_2H_4$ and the regions where all Fe positions in the row are visible. This indicates that all protrusions are probably due to $C_2H_4$. The molecules are pinned at defects, and the neighbors of a pinned molecule will be usually at a distance of 6 Å along the row. With increasing distance from the defects, the molecules are increasingly mobile, at a timescale shorter than that of imaging by the STM (note that mobility may be also aided by the STM tip). It is also possible that the tip is terminated by an ethylene molecule, which may facilitate the high resolution observed in Fig. 4c.

The measurement temperature of 78 K falls close to the onset of the TPD desorption peak δ. Since the timescale of STM measurements (many minutes) is longer than that of TPD, the coverage seen by STM will be lower than that in TPD at the same temperature. Therefore, we pumped the cryogen in the cryostat of the STM to reduce the sample temperature to ≈68 K, which should allow to saturate the δ peak. Following exposure to $C_2H_4$ (an additional 2.3 L was dosed in addition to that shown in Fig. 4c), we observe that the surface Fe rows are no longer visible, and the surface is imaged as a complete layer of circular protrusions (Fig. 4d). As the molecules on the Fe rows are packed as closely as possible, we find indications of mobility in very few places only (Fig. 4d red arrow).

Some of the molecules appear clearly brighter than others. Images acquired before and after exposure to ethylene on the same sample area, as well as a quantitative analysis of the number of bright molecules and surface defect concentration lead us to the conclusion that the brighter protrusions are due to ethylene adsorbed on surface defects. More details of this analysis can be found in the supporting information (see Fig. S5). Further cooling of the sample (i.e., using liquid He as the cryogen), to stabilize higher coverages, is hampered because the $Fe_3O_4(001)$ sample is not conductive enough for STM measurements at 4 K.

### 3.4 DFT Calculations

A systematic approach was utilized to determine the lowest-energy configurations of the ethylene molecules at relevant coverages on the $Fe_3O_4(001)$ surface. We utilized the strongly constrained and appropriately normed meta-generalized gradient approximation (SCAN) with the inclusion of van der Waals interactions (rVV10), but a comparison to several alternatives (generalized gradient-based functionals, GGA) with and without van der Waals corrections is included at the end of this section. All other settings of the DFT calculations and convergence criteria are the same for all functionals. The selection of the SCAN functional over PBE was motivated by its demonstrated superior performance on various molecular and solid-state test sets, as reported in the literature[34, 50]. Moreover, SCAN is an attractive choice as it is reported to balance accuracy and computational efficiency[51].

The SCV termination of the clean $Fe_3O_4(001)$ surface has four equivalent five-fold undercoordinated Fe atoms ($Fe_{oct}$) (truncated octahedral coordination, thus named $Fe_{oct}$) and eight surface O atoms per $(\sqrt{2} \times \sqrt{2})R45°$ unit cell (two thereof are twofold coordinated, and six threefold). We utilized a $(2\sqrt{2} \times 2\sqrt{2})R45°$ supercell to explore many different configurations in the coverage regime 1–4 $C_2H_4$/unit cell. We also modelled three experimentally observed defects in this surface[46]: (1) an Fe adatom, (2) an unreconstructed unit cell (i.e., an additional Fe atom in the third layer that allows to locally recover the spinel structure), and (3) a surface hydroxy group. For coverages of more than one molecule per unit cell, the calculated adsorption energy reported here is the average adsorption energy per $C_2H_4$ molecule (see equation (1)).

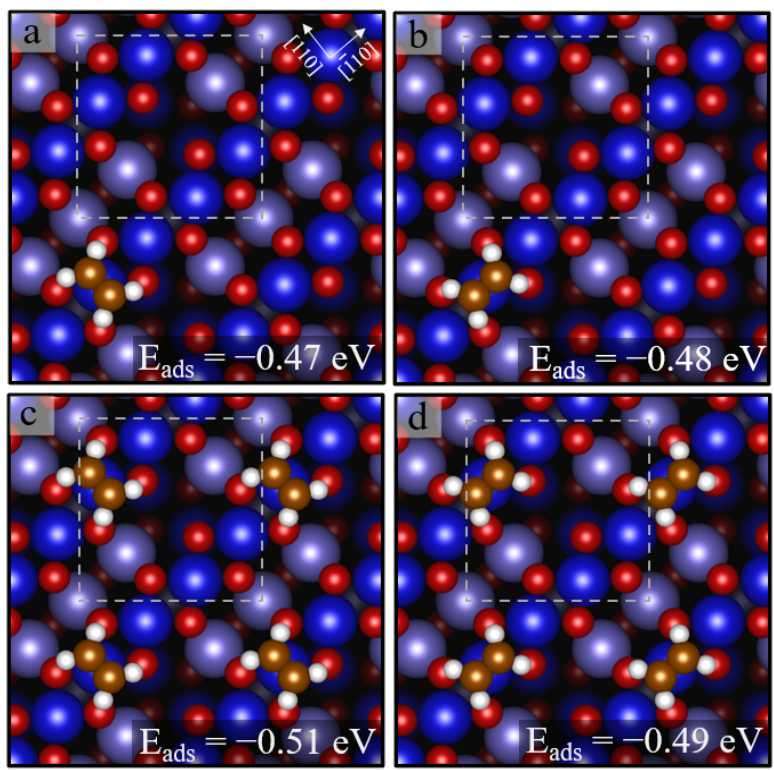

**Figure 5. Low-coverage structures (up to one $C_2H_4$ molecule per unit cell) determined by DFT+U (SCAN+rVV10) (top view).** (a) and (b) for isolated $C_2H_4$ molecules and (c) and (d) for one $C_2H_4$/unit cell. The white, dashed square indicates the $(\sqrt{2}\times\sqrt{2})R45°$ unit cell. Surface $Fe_{oct}$ atoms are dark blue, tetrahedrally coordinated Fe atoms ($\approx 0.8$ Å lower than the $Fe_{oct}$) slate blue, and oxygen is red.

On the defect-free $Fe_3O_4$(001) surface, we find that an isolated $C_2H_4$ molecule adsorbs molecularly on top of the five-fold $Fe_{oct}$ atom and adopts a flat-lying geometry with the C=C bond along or perpendicular to the Fe row ($E_{ads}$ = −0.48 eV or −0.47 eV, respectively). These configurations are similar to those calculated for the $RuO_2$(110) surface[52], where easy in-plane rotational motion of the π-$C_2H_4$ complex was found at the low coverage. Bonding via the π orbital to an undercoordinated cation was also found for $C_2H_4$ on rutile $TiO_2$(110)[53]. Increasing the coverage to one $C_2H_4$ per unit cell results in a slightly stronger adsorption energy ($E_{ads}$= −0.51 parallel to the Fe row, −0.49 eV perpendicular, see Figure 5c and 5d). We attribute this to van der Waals attraction between neighboring molecules.

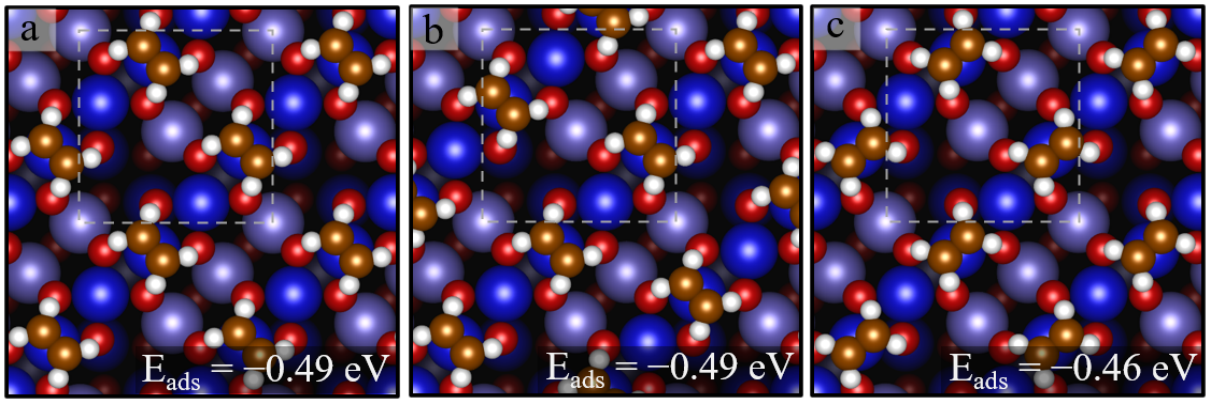

**Figure 6. Structures for two C$_2$H$_4$ molecules per unit cell determined by DFT+U (top view).** (a) and (b) C$_2$H$_4$ molecules adsorb on top of Fe$_{oct}$ atoms, with the C=C bond aligned perpendicular to the row, and (c) C$_2$H$_4$ molecules adsorb on top of Fe$_{oct}$ atoms, with the C=C bond aligned along the row. The white dashed square indicates the ($\sqrt{2}\times\sqrt{2}$)R45° unit cell.

At a coverage of two C$_2$H$_4$ molecules per unit cell, the corresponding adsorption energy is calculated as −0.49 eV in the configurations of Figs. 6a and 6b. The optimal orientation of the C=C bond is perpendicular to the Fe$_{oct}$ row. Aligning the C=C bond along the Fe$_{oct}$ row yields a weaker adsorption energy of −0.46 eV (Fig. 6c).

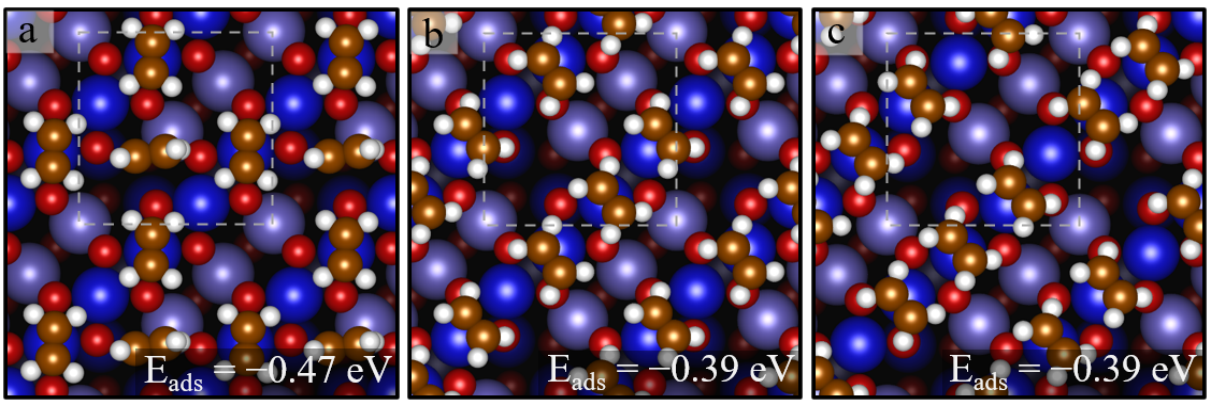

**Figure 7. Structures for three C$_2$H$_4$ molecules per unit cell determined by DFT+U (top view).** (a) two C$_2$H$_4$ molecules adsorb on the Fe$_{oct}$ row, and one C$_2$H$_4$ molecule adsorbs between the Fe$_{oct}$ rows, (b) and (c) three C$_2$H$_4$ adsorb on the Fe$_{oct}$ row.

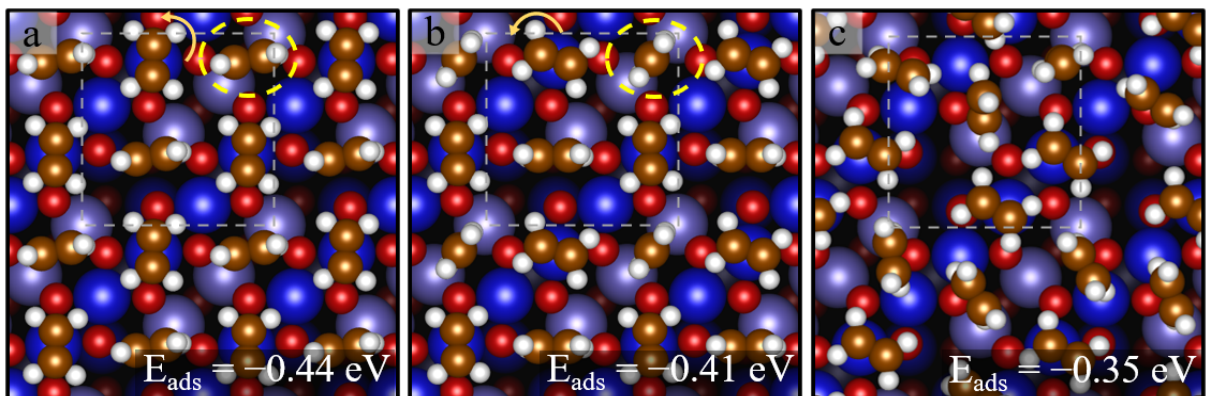

**Figure 8: Structures for four C$_2$H$_4$ molecules per unit cell determined by DFT+U (top view).** (a) two C$_2$H$_4$ molecules adsorb on the Fe$_{oct}$ row, and the other two C$_2$H$_4$ molecules adsorb between the Fe$_{oct}$ rows where the two ethylene molecules lie flat on the surface. (b) two C$_2$H$_4$ molecules adsorb on the Fe$_{oct}$ row, and the other two C$_2$H$_4$ molecules adsorb between the Fe$_{oct}$ rows where one ethylene molecule is positioned upright on the surface, shown in a yellow dashed oval. (c) an attempted configuration where the C$_2$H$_4$ molecules were initially placed on every Fe$_{oct}$ atom on the row. The adsorbed C$_2$H$_4$ molecules to tilt or move away from the ideal-on-top of five-fold Fe$_{oct}$ atoms during the structure optimization. The adsorption is weaker by 0.06–0.08 per ethylene molecule compared with the (a) and (b) configurations. The white dashed square indicates the ($\sqrt{2}\times\sqrt{2}$)R45° unit cell.

Our calculations (Fig. 7) clearly show that placing three C$_2$H$_4$ on neighboring surface Fe$_{oct}$ sites is substantially less favorable than the structures at 1–2 C$_2$H$_4$ per cell. When placing more than two molecules per cell on the Fe rows, repulsive interactions between neighboring molecules cause the C$_2$H$_4$ molecules to tilt away from the ideal atop geometry, leading to a significant weakening of the average adsorption energy (Figs. 7b and 7c). Our DFT results show that the better option is half occupation of the Fe rows and placing the additional C$_2$H$_4$ molecule between the Fe$_{oct}$ rows (Fig. 7a). The minimum-energy structure of Fig. 7(a) exhibits a motif of the crystal structure of solid C$_2$H$_4$ [57]: The H atoms of the "additional" molecules in the trough point towards the C=C bonds of the molecules on the Fe rows.

Searching for the optimal configuration at four molecules per unit cell is complicated by the many possibilities, but it is clear that two molecules adsorb atop Fe$_{oct}$ atoms, with the other two molecules adsorbed weakly in between the rows. Two possibilities are shown in Fig. 8a and 8b. A configuration with initially four C$_2$H$_4$ molecules all adsorbed atop Fe$_{oct}$ sites is not stable; the molecules all moved away from the ideal geometry after structural optimization (Fig. 8c) and the final structure was still less favorable than the other models. Overall, we can conclude that the coverage of two molecules per unit cell is a threshold above which additional molecules must be placed in relatively unfavorable configurations between the surface Fe rows. This nicely explains why the δ TPD peak saturates at a coverage of 2 C$_2$H$_4$ per unit cell.

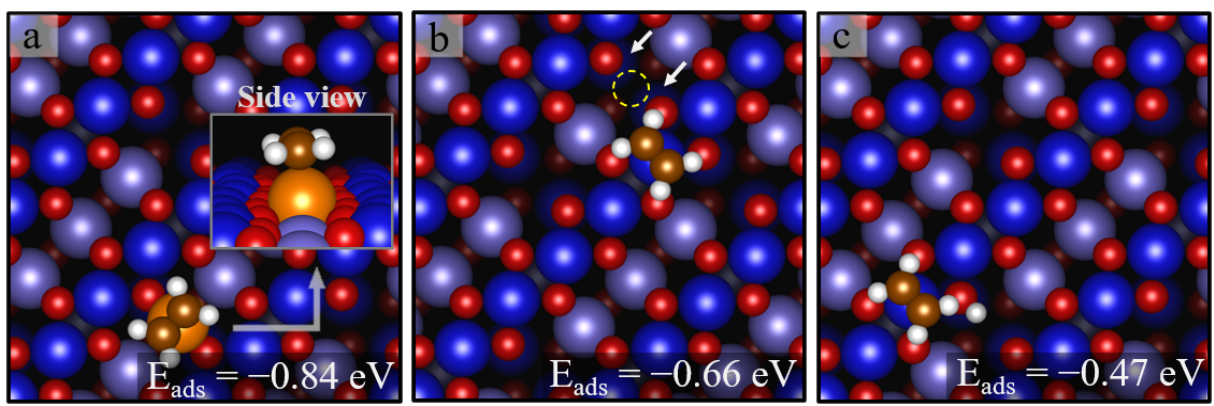

**Figure 9. Structures for defects determined by DFT+U (top view).** (a) an Fe adatom (orange) between the Fe$_{oct}$ rows, in (b) an unreconstructed unit cell (i.e., two additional Fe atoms in the third layer at the position marked by white arrows. Adding these atoms and removing a tetrahedral Fe$_{int}$ atom from the site marked by a yellow dashed circle locally recover the spinel structure), and in (c) a surface hydroxy group is shown.

We also computed the C 1s core-level binding energies in the final state approximation. The relative shift in binding energy between carbon atoms contained within C$_2$H$_4$ adsorbed on top of the Fe$_{oct}$ atoms is 0.3 eV lower that those located between the rows at high coverage. This result is qualitatively consistent with our XPS data. The results were not very sensitive to the exact configuration of the additional molecule, because the molecule between the Fe$_{oct}$ rows is bound mainly by van der Waals forces and its electrostatic quadrupole moment.

Finally, we turn our attention to adsorption at the surface defects. We find that C$_2$H$_4$ adsorbed at an Fe adatom defect exhibits the strongest adsorption energy of all sites considered here (−0.84 eV). The optimal configuration has the molecule atop the adatom, and the C=C parallel to the surface Fe rows (Fig. 9a). The second type of defect (Fig. 9b) considered is the so-called "unreconstructed unit cell", in which the second layer Fe$_{int}$ atom is replaced by two third-layer Fe$_{oct}$ atoms. This can be viewed as a local recovery of the bulk spinel structure. C$_2$H$_4$ binds on this surface with an energy of −0.66 eV. Since this defect is locally similar to the APDB[46], we assume that adsorption at the APDB would yield a similar adsorption energy. Attempts to bind C$_2$H$_4$ adjacent to a hydroxy group resulted in a relatively weak adsorption energy (E$_{ads}$ = −0.47 eV), as shown in Fig. 9c. Formation of a C$_2$H$_5$ radical (by using the H atom from the hydroxy group) is less favorable than the separate hydroxyl and adsorbed C$_2$H$_4$ by 0.8 eV. This is substantially less than the adsorption energy of C$_2$H$_4$. We conclude that the hydroxyl group does not infer any significant change in the C$_2$H$_4$ adsorption energy (to within the uncertainty).

We found that the calculated $E_{ads,a}$ value are quite sensitive to the DFT approach chosen. Table 1 lists the calculated C$_2$H$_4$ adsorption energies at a coverage of 1 and 2 C$_2$H$_4$ per unit cell using different functionals that take account of van der Waals (vdW) effects as well as two common functionals without vdW corrections (PBE, SCAN).

**Table 1: Comparison of the average adsorption energies obtained for 1 and 2 $C_2H_4$ on $Fe_3O_4(001)$ using various functionals.**

| Coverages | PBE | PBE-D2[54] | PBE-D3[55] | optPBE-DF[56] | optB86-DF[57] | optB88-DF[56] | SCAN[33] | SCAN+rVV10[34] |
|---|---|---|---|---|---|---|---|---|
| 1 $C_2H_4$ | −0.19 | −0.50 | −0.54 | −0.50 | −0.53 | −0.58 | −0.38 | −0.51 |
| 2 $C_2H_4$ | −0.19 | −0.48 | −0.50 | −0.52 | −0.56 | −0.57 | −0.35 | −0.49 |

Including van der Waals functionals, both semi-local and nonlocal, leads to higher adsorption energies than found experimentally. All functionals yield adsorption energies with a very weak coverage dependence for 1 and 2 $C_2H_4$ per unit cell. Among the vdW-corrected functionals, the D2 and SCAN+rVV10 provide the closest match to the experimental value. Table 1 shows that the vdW correction in the SCAN level is smaller than in the PBE level. This difference can be attributed to the fact that SCAN is capable of capturing intermediate-range London dispersion interactions, which allows it to capture noncovalent interactions more accurately, ultimately resulting in smaller vdW corrections for long-range interactions compared to PBE.

Including van der Waals functionals, both semi-local and nonlocal, leads to higher adsorption energies than found experimentally. All functionals yield adsorption energies with a very weak coverage dependence for 1 and 2 $C_2H_4$ per unit cell. Among the vdW-corrected functionals, the D2 and SCAN+rVV10 provide the closest match to the experimental value. Table 1 shows that the vdW correction of SCAN+rVV10 is smaller than for the various vdW-corrected PBE functionals. This difference can be attributed to the fact that the SCAN functional alone (without vdW corrections) is capable of capturing intermediate-range London dispersion interactions[33], which allows it to describe noncovalent interactions more accurately, ultimately resulting in smaller vdW corrections for long-range interactions compared to PBE.

Table 2 shows a full summary of the experimental data from TPD, the differential adsorption energies gained from the TPD analysis and the adsorption energy calculated using DFT. As already mentioned earlier in this section, the differential adsorption energy is the adsorption energy of the most weakly bound $C_2H_4$ molecule in a structure ($E_{ads,d}$, see equation (2)). The values calculated by DFT+U (SCAN+rVV10) indicate stronger binding by approximately 0.1–0.15 eV, than the value from the TPD analysis. We attribute the difference to the limited accuracy of the vdW functionals. A comparison of the TPD differential adsorption energies with the desorption energies from Ref. 25 where a different analysis method (inversion analysis) was used, shows very good agreement for the δ and ε peaks. The defect peak η is not as well defined as the others and thus its desorption energy has a rather large error bar. It should be noted that the sign differs because the analysis of Ref. 25 provides desorption barriers (positive) whereas our analysis yields adsorption energies, but the absolute values should be approximately the same since the difference is mainly the adsorption barrier, which should be negligible in the present case (near-unity sticking, see Fig. S1).

**Table 2: Summary of the experimental and computational results for the desorption peaks and those determined experimentally in Ref. 25.**

| TPD Peaks | Description | $C_2H_4$ / unit cell saturation | Desorption Temperature (K) | TPD-Analysis $E_{ads,d}$ (eV) | $E_{ads,d}$ from DFT (SCAN+rVV10) (eV) | Inversion analysis in Ref. 25 $E_{des,d}$ (eV) |
|---|---|---|---|---|---|---|
| α | Multilayer | 12.7 (±1.2) | 73-76 | - | - | - |
| β | Multilayer | 8.9 (±0.8) | 70-73 | - | - | - |
| γ | Saturated monolayer | 4.1 (±0.4) | 90 | −0.23 | −0.35 | - |
| δ | ¾ monolayer | 3.0 (±0.3) | 100 | −0.32 | −0.43 | 0.34 |
| ε | ½ /¼ monolayer | 2.2 (±0.2) | 110-115 | −0.36/−0.40 | −0.47 / −0.51 | 0.37/0.42 |
| η | Defects | ≈0.3 | 160 | −0.54/≈−0.6 | −0.66 / −0.84 | ≈ 0.51 |

## 4. Discussion

Combining the information garnered from the various techniques employed in this work, it is possible to build up a comprehensive picture of how $C_2H_4$ interacts with the $Fe_3O_4$(001) surface. $C_2H_4$ binds most strongly at surface defects such as Fe adatoms, unreconstructed unit cells and APDB's, as seen in the STM image in Fig. 4b. Desorption from these sites results in the η peak in TPD and may also contribute some intensity in the region down to the ζ peak. Our DFT calculations suggest that the Fe adatoms bind strongest of all, followed by unreconstructed unit cells. Since the APDB is locally similar in structure to an unreconstructed unit cell, we assume similar adsorption properties, even though we did not calculate an APDB explicitly. Surface OH groups are omnipresent on $Fe_3O_4$(001), but there is no evidence that they interact strongly with $C_2H_4$.

As mentioned above, and as noted by Lee et al.,[25] the ζ TPD peak exhibits unusual behavior, as it continues to grow following the onset of the ε peak. This suggests that diffusion cannot occur between the sites responsible for these desorption peaks. Interestingly, we have observed that this peak is almost absent on a brand new $Fe_3O_4$(001) sample (as seen in Fig. 1), but increases in intensity as the crystal is utilized for experiments (see Fig. S2). This behavior suggests that its origin may lie in the $Fe_2O_3$ inclusions that slowly grow over time as an $Fe_3O_4$(001) sample is oxidized during UHV preparation[26]. As such, we do not consider this peak as representative of the Fe$_3$O$_4$(001) surface. It should be also noted that the temperature range with the largest discrepancy between the experimental spectra and simulation in Fig. 2 is around the ζ peak. Since the simulation is based on equilibrium thermodynamics, it does not correctly describe the case of widely separated different regions on the surface that cannot quickly equilibrate through diffusion. In addition, in the temperature range of the ζ peak (120 K), magnetite bulk undergoes a phase transition (the so-called Verwey transition[58]), which can be also observed at the surface[59]. This transition substantially changes the electronic structure of the sample, which could affect the adsorption energy of adsorbed molecules. Thus, part of the unusual behavior in TPD around this temperature may also be related to the Verwey transition.

The ε TPD peak emerges at ≈0.3 C$_2$H$_4$ per unit cell, and shifts slightly to lower temperature before saturating at approximately 2 C$_2$H$_4$ per unit cell (Fig. 1). This is likely due to repulsion between the C$_2$H$_4$ molecules. The DFT+U results show slight weakening of the adsorption energy shown between 1 and 2 C$_2$H$_4$ per unit cell in Figs. 5 ($E_{ads}$ = −0.51 eV) and 6 ($E_{ads}$ = −0.49 eV), respectively, but this difference is certainly within the error of the calculation.

Calculations performed using other functionals (see Table 1) also suggest that the adsorption energies at 1 and 2 molecules per unit cell are very close, i.e., their interaction is weak. Quantitatively, however, we note that all the tested functionals with van der Waals corrections significantly overestimate the adsorption strength at 2 C$_2$H$_4$ per unit cell compared to the experimentally determined average adsorption energy of −0.38 eV. The overestimation is largest in the case of optB88 (0.19 eV), which is in line with the 0.2–0.3 eV overestimation observed for CO adsorbed in various Fe$_3$O$_4$-based SAC systems[60]. With the present weak binding, however, this would result in the adsorption being too strong by about 50%. In retrospect, it is perhaps unsurprising that including vdW functionals such as optB88 do not perform quantitatively well for metal oxide systems, because they are typically optimized to account for the van der Waals interactions between gas-phase molecules and not for the interaction of molecules with surfaces.

Based on the STM, DFT and TPD results, we conclude that the C$_2$H$_4$ molecules prefer to occupy the next-nearest neighbor positions along the surface Fe rows on the defect-free surface. Fig. 4c indicates that molecules are pinned at surface defects, but otherwise mobile even at 78 K. The STM image shows a time average of the positions of C$_2$H$_4$ on the surface, which is why the apparent height of the molecules slowly decreases away from the defect until the protrusions appear equally bright at all Fe atoms at a greater distance. When the sample was cooled further it became possible to complete the next nearest neighbor periodicity (6 Å) along the surface Fe rows, but the overall coverage cannot be determined from the STM images. On the one hand it could be two C$_2$H$_4$ per unit cell, as it appears based on the density of protrusions in Fig. 4d, but it could also be that molecules are adsorbed between the rows but are not directly imaged. In either case, the preference for next nearest neighbor site occupancy on the Fe rows is clear.

Next, we turn to the δ and γ peaks, which occur between 2 $C_2H_4$ per unit cell and saturation of the monolayer at ≈4 $C_2H_4$ per unit cell. These peaks are particularly sharp, and are not well reproduced by the TPD simulation in Fig. 2b. It is important to note that the TPD simulation program[40] does not include intermolecular interactions. Repulsive reactions are approximated as an occupation of sites with lower desorption energy as the coverage is increased (as in the case of the ε peak). Attractive interactions, on the other hand, can be inferred when the TPD peak is sharper than the peak width predicted based on occupation of a site with a singular desorption energy. This is in line with the DFT model shown in Fig. 7a, for example, where the molecules orient to maximize attraction in a similar fashion to that seen in crystalline $C_2H_4$.

Finally, we note some interesting behavior within the $C_2H_4$ multilayer. Once saturation of the first layer is completed at 4 $C_2H_4$/ unit cell, additional molecules begin to desorb from a peak at lower temperature. This peak has a zero-order line shape typical for multilayer desorption. Once the coverage reaches ≈8 $C_2H_4$ per unit cell, the leading edge (and with it the rest of the peak) shifts in its entirety to a higher temperature. This suggests that the structure changes abruptly once the 2$^{nd}$ layer is completed and the 3$^{rd}$ layer begins to grow. This might occur because the structure of the two-layer system is defined by the structure of the first monolayer, while the system adopts a more stable 3-dimensional crystal structure once additional molecules adsorb on top.

In closing, we remark that the similarity of the experimental TPD data between our work and that of Lee et al.[25] demonstrates that $Fe_3O_4(001)$ is a well-reproducible model system, and thus suitable for surface science investigations of iron-oxide surface chemistry. The TPD peaks appear at very similar temperatures, so the differences in the desorption energy calculated (see Table 2) is a result of different assumptions made in the analysis of the TPD data. As mentioned by Lee et al. in ref. 25, the inversion analysis method is not well suited to broad desorption spectra such as those found for $C_2H_4$ on $Fe_3O_4(001)$ because it is most sensitive to the leading edge. Our method utilizes DFT to estimate the vibrational entropy of the adsorbed molecules, and thus does not rely on the determination of a desorption prefactor from the experimental data. At coverages of 2 and 3 $C_2H_4$/unit cell, where the peaks are relatively sharp, the difference is approximately 0.02 eV, which is certainly less than the error of the methods. In any case, the weak, non-dissociative binding of $C_2H_4$ along with the ability of $Fe_3O_4(001)$ to stabilize dense arrays of metal adatoms further makes this an ideal candidate model system to investigate hydrogenation and hydroformylation reactions by metal oxide-supported single atoms.

## 5. Conclusion

We investigated the interaction of a representative alkene, $C_2H_4$, with the $Fe_3O_4(001)$ surface by utilizing a combination of different techniques; TPD, XPS, STM and DFT computations. $C_2H_4$ adsorbs most strongly at surface defects where it desorbs at ≈160 K. DFT-based calculations predict an adsorption energy of −0.84 to −0.66 eV depending on the defect site. A broad desorption peak appears at 110-115 K and saturates at approximately 2 $C_2H_4$ per unit cell. The differential adsorption energies of 1 and 2 $C_2H_4$ molecules per unit cell were calculated by DFT as −0.51 and −0.47 eV, respectively, while TPD yields −0.40 and −0.36 eV. Some vdW-corrected DFT functionals overestimate the adsorption energy even more (by up to 0.18 eV). STM and DFT results suggests that up to a coverage of 2 $C_2H_4$ per unit cell the molecules prefer to occupy the next-nearest neighbor positions along the surface Fe rows. Suggested by DFT, additional $C_2H_4$ are stabilized through attractive intermolecular interactions between the Fe rows up to a coverage 4 $C_2H_4$ molecules per unit cell. Multilayer desorption happens at 73–76 K. Our study shows that $C_2H_4$ weakly adsorbs at the $Fe_3O_4(001)$ surface and desorbs molecularly below room temperature with a monolayer coverage of 4 $C_2H_4$ molecules per unit cell.

**Supporting Information**.

The supporting information for this manuscript contains:

List of initial coverages for the TPD series shown in Fig. 1.
Sticking co-efficient measurement for $C_2H_4$ on $Fe_3O_4(001)$.
TPD data from a second sample showing the effect of sample aging.
XPS data for $C_2H_4$ on $Fe_3O_4(001)$ (O1s, Fe2p, C1s).
STM images of the $Fe_3O_4(001)$ surface during exposure to ethylene.

**Acknowledgements**

LP, GSP, JP, PS, ARA and MM acknowledge funding from the European Research Council (ERC) under the European Union's Horizon 2020 research and innovation programme (grant agreement No. [864628], Consolidator Research Grant 'E-SAC'). This work was also supported by the Austrian Science Fund (FWF) under project number F81, Taming Complexity in Materials Modeling (TACO) (MM, CW, MU, JP, GSP, UD, MSchmid, CF). The computational results presented have been achieved by using the Vienna Scientific Cluster.

Table of contents graphic:

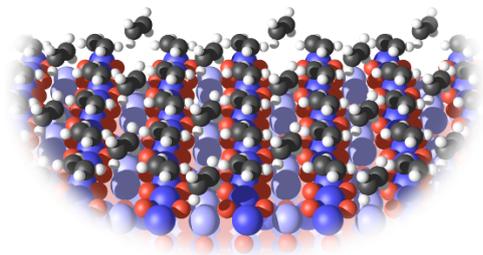

# Supporting Information for
# A Multi-Technique Study of $C_2H_4$ Adsorption on $Fe_3O_4$(001)

Lena Puntscher[1*], Panukorn Sombut[1*], Chunlei Wang[1], Manuel Ulreich[1], Jiri Pavelec[1], Ali Rafsanjani-Abbasi[1], Matthias Meier[1,2], Adam Lagin[1], Martin Setvin[1,3], Ulrike Diebold[1], Cesare Franchini[2,4], Michael Schmid[1] and Gareth S. Parkinson[1]

[1]Institute of Applied Physics, TU Wien, Vienna, Austria
[2]Faculty of Physics, Center for Computational Materials Science, University of Vienna, Vienna, Austria
[3]Department of Surface and Plasma Science, Faculty of Mathematics and Physics, Charles University, Prague, Czech Republic
[4]Dipartimento di Fisica e Astronomia, Università di Bologna, Bologna, Italy

**Initial coverages for the TPD series shown in Fig. 1:**
Figure 1 shows a series of TPD spectra for various initial coverages of (0.3, 0.6, 0.9, 1.2, 1.6, 2.1, 2.4, 2.7, 2.9, 3.0, 3.4, 3.9, 4.2, 4.6, 4.9, 5.8, 6.0, 6.7, 7.5, 8.4, 9.1, 12.1) $C_2H_4$ molecules per $(\sqrt{2}\times\sqrt{2})R45°$ unit cell.

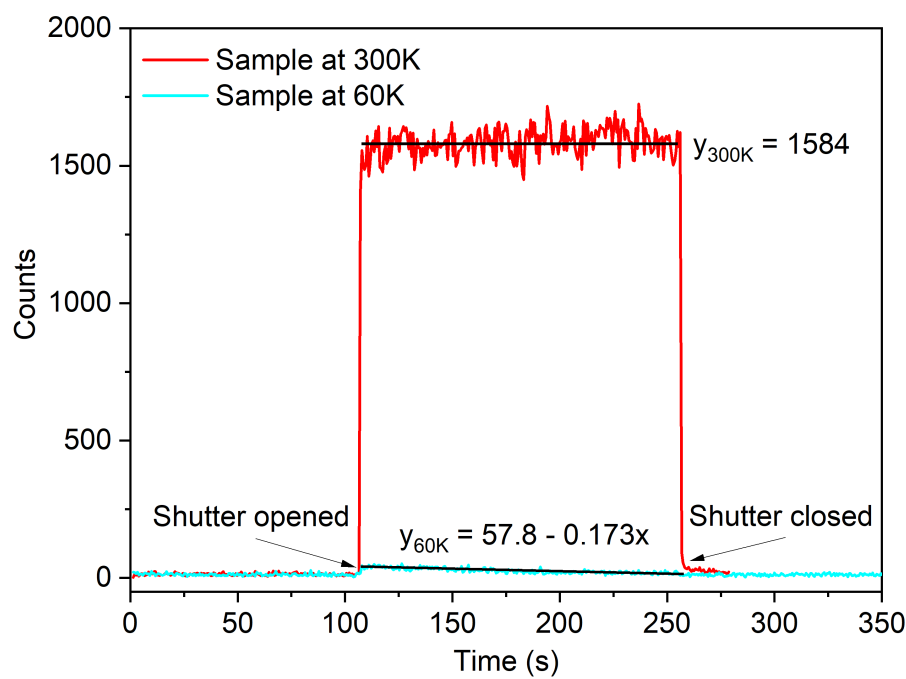

**Figure S1:** The sticking co-efficient measurement performed here is based on the King and Wells approach[1]. For the reference of zero sticking, the as-prepared sample is held in UHV at 300 K, which is well above the temperature of any peaks observed in TPD (see Figure 1 in the main text). After a time 105 s, the molecular beam shutter is opened and $C_2H_4$ molecules impinge on the sample at normal incidence. Molecules reflect from the surface and scatter into the vacuum system, and some are measured by the mass spectrometer. Ideally, the mass spectrometer is positioned in a non line of sight geometry to prevent direct scattering into the mass spectrometer. In our setup, however, this is not possible as the angle between the mass spectrometer and molecular beam source is fixed at 30°. When this experiment is repeated at 60 K, the signal is much lower because most of the molecules are adsorbed at the sample surface. To determine the sticking coefficient, one has to calculate the difference between the two curves as a function of time and divide it by the background-corrected count rate at zero sticking, $y_{300K}$. In our measurement shown in Fig. S1, the signal acquired for the 300 K measurement is constant, as expected. The 60 K signal begins at 2.5% of the intensity (after subtraction of the background) and decreases approximately linearly to 0 (i.e. 100% sticking) after 150 seconds. The slow increase to 100% sticking is typical for such molecules on surfaces, and occurs because momentum transfer is maximised once molecules arrive at a surface already covered in similar molecules. The formulae shown in the figure result from a linear fit to the data for the duration of time that the shutter was open. A more thorough description of sticking coefficient measurements is contained within the work of Chen et al.[2].

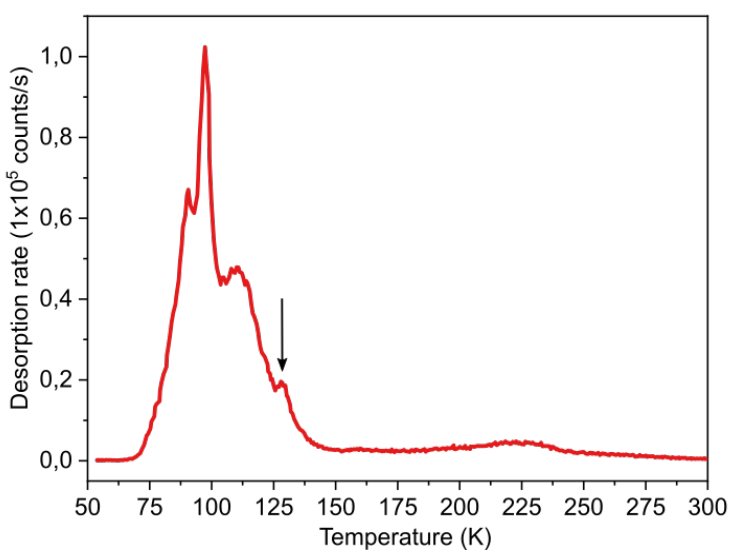

**Figure S2:** TPD curve for 4.1 $C_2H_4$/unit cell acquired on a sample previously utilized daily for surface science experiments in our setup for approximately 1 year. The data is similar to that obtained on a freshly installed sample (see Figure 1 in the main text), apart from an enhanced intensity of the shoulder at 125 K, which becomes a peak (see black arrow). This peak was previously observed with a similar intensity by Lee et al.[3]. One possible explanation for this peak is adsorption at $\alpha$-$Fe_2O_3$ inclusions. De la Figuera and coworkers have shown that a typical annealing cycle with a partial pressure of $10^{-6}$ mbar $O_2$ leads to the growth of many hundreds of layers of virgin $Fe_3O_4$(001) surface [4,5]. This is one of the reasons why this surface is relatively straightforward to prepare, but the iron required for this growth is obtained by oxidizing the sample overall. Rather than a homogeneous distribution of iron vacancies, the oxidation manifests in the growth of $\alpha$-$Fe_2O_3$ inclusions, which grow along the <110> directions at the $Fe_3O_4$(001) surface. In extreme cases, these can be visible to the eye as a chequerboard appearance on the sample surface, as can be seen in Ref. [5]. In the early stages it appears more as a matte appearance of the sample surface, compared the extremely polished appearance of the as-purchased samples.

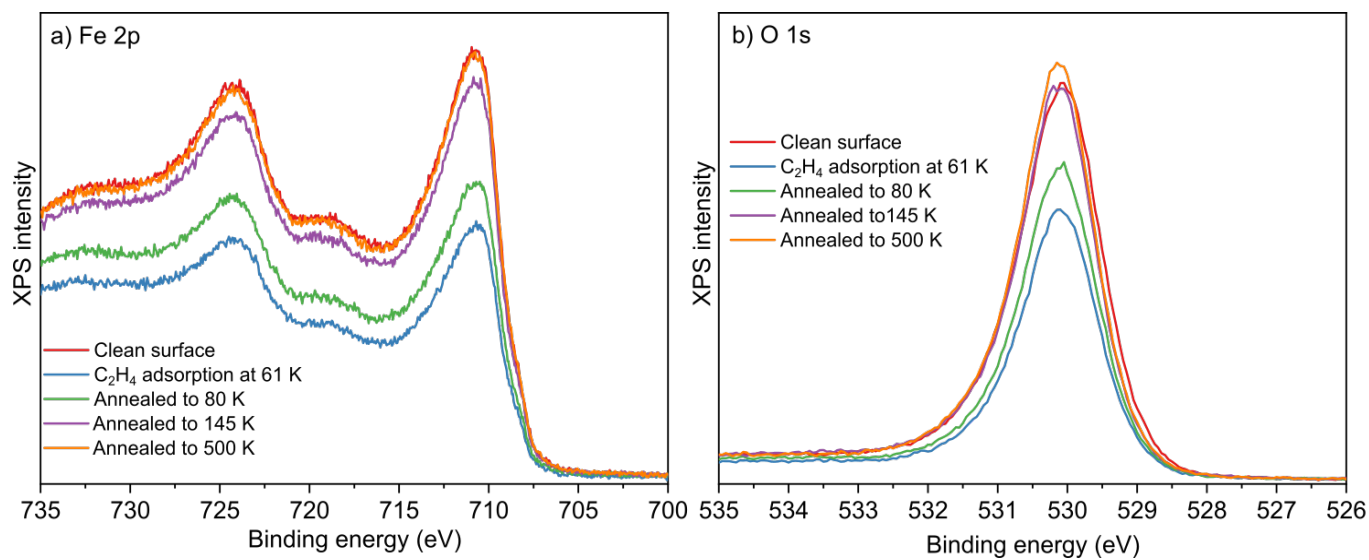

**Figure S3:** X-ray photoelectron spectroscopy data for the $C_2H_4/Fe_3O_4(001)$ system measured at 61 K after $C_2H_4$ adsorption and after several heating steps up to 500 K. In a) the Fe2p region is shown, in b) the O1s region.

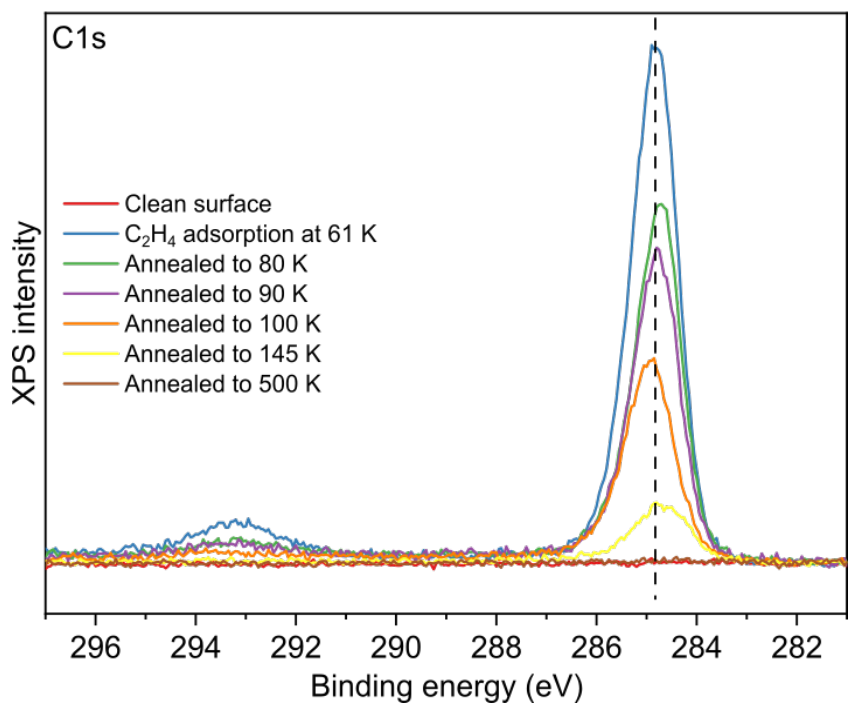

**Figure S4:** X-ray photoelectron spectroscopy data for the $C_2H_4/Fe_3O_4(001)$ system measured at 61 K after $C_2H_4$ adsorption and after several heating steps up to 500 K. A second peak 8.3 eV shifted to higher binding energy from the C1s peak at 284.8 eV is due to a pi-3p Rydberg shake-up process.

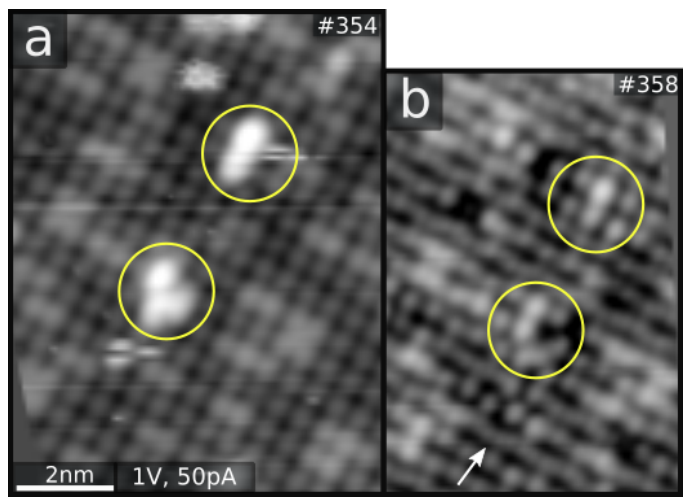

**Figure S5:** STM images of the $Fe_3O_4$(001) surface during exposure to ethylene. In (a) ethylene molecules only adsorb at defect sites, while in (b) also the $Fe^{3+}$ rows start to get occupied. There are still $Fe^{3+}$ rows that appear unoccupied (marked with an arrow). The yellow circles show a defect site which is occupied by ethylene in both images. It is clearly visible in (b) that the ethylene molecules adsorbed at defect sites appear brighter than the ethylene molecules adsorbed at regular lattice sites.